\documentclass[conference]{IEEEtran}
\IEEEoverridecommandlockouts
\usepackage{cite}
\usepackage{amsmath,amssymb,amsfonts}
\usepackage{algorithmic}
\usepackage{graphicx}
\usepackage{textcomp}
\usepackage{xcolor}
\usepackage[ruled, linesnumbered, vlined]{algorithm2e}

\def\BibTeX{{\rm B\kern-.05em{\sc i\kern-.025em b}\kern-.08em
    T\kern-.1667em\lower.7ex\hbox{E}\kern-.125emX}}
\begin{document}

\title{Cargo Ecosystem Dependency-Vulnerability Knowledge Graph Construction and Vulnerability Propagation Study}

\author{
\IEEEauthorblockN{Peiyang Jia}
\IEEEauthorblockA{Xidian University, 
  China \\National Computer Network and \\ Intrusion Protection Center \\ 
  University of Chinese \\ Academy of Sciences ,China\\
jiapy@nipc.org.cn}
\and

\IEEEauthorblockN{Chengwei Liu}
\IEEEauthorblockA{Nanyang Technological University\\
chengwei001@e.ntu.edu.sg}
\and
\IEEEauthorblockN{Hongyu Sun}
\IEEEauthorblockA{Xidian University, China\\
National Computer Network Intrusion Protection Center\\ University of Chinese Academy of Sciences ,China\\
sunhy@nipc.org.cn}
\and

\IEEEauthorblockN{Chengyi Sun}
\IEEEauthorblockA{National Computer Network Intrusion Protection Center\\ University of Chinese Academy of Sciences ,China\\
suncy@nipc.org.cn}
\and

\IEEEauthorblockN{Mianxue Gu}
\IEEEauthorblockA{Hainan University,China\\
National Computer Network Intrusion Protection Center\\ University of Chinese Academy of Sciences ,China\\
gumx@nipc.org.cn}
\and

\IEEEauthorblockN{Yang Liu}
\IEEEauthorblockA{Nanyang Technological University\\
yangliu@ntu.edu.sg}
\and

\IEEEauthorblockN{Gaofei Wu}
\IEEEauthorblockA{Xidian University, China\\
National Computer Network Intrusion Protection Center\\ University of Chinese Academy of Sciences ,China\\
gfwu@xidian.edu.cn}
\and

\IEEEauthorblockN{He Wang}
\IEEEauthorblockA{Xidian University, China\\
National Computer Network Intrusion Protection Center\\ University of Chinese Academy of Sciences ,China\\
hewang@xidian.edu.cn}
\and

\IEEEauthorblockN{Yuqing Zhang}
\IEEEauthorblockA{Xidian University, China\\
Hainan University\\
National Computer Network Intrusion Protection Center\\ University of Chinese Academy of Sciences ,China\\
zhangyq@nipc.org.cn}
\and

}

\maketitle


\begin{abstract}
Currently, little is known about the structure of the Cargo ecosystem and the potential for vulnerability propagation. Many empirical studies generalize third-party dependency governance strategies from a single software ecosystem to other ecosystems but ignore the differences in the technical structures of different software ecosystems, making it difficult to directly generalize security governance strategies from other ecosystems to the Cargo ecosystem. To fill the gap in this area, this paper constructs a knowledge graph of dependency vulnerabilities for the Cargo ecosystem using techniques related to knowledge graphs to address this challenge. This paper is the first large-scale empirical study in a related research area to address vulnerability propagation in the Cargo ecosystem. This paper proposes a dependency-vulnerability knowledge graph parsing algorithm to determine the vulnerability propagation path and propagation range and empirically studies the characteristics of vulnerabilities in the Cargo ecosystem, the propagation range, and the factors that cause vulnerability propagation. Our research has found that the Cargo ecosystem's security vulnerabilities are primarily memory-related. 18\% of the libraries affected by the vulnerability is still affected by the vulnerability in the latest version of the library. The number of versions affected by the propagation of the vulnerabilities is 19.78\% in the entire Cargo ecosystem. This paper looks at the characteristics and propagation factors triggering vulnerabilities in the Cargo ecosystem. It provides some practical resolution strategies for administrators of the Cargo community, developers who use Cargo to manage third-party libraries, and library owners. This paper provides new ideas for improving the overall security of the Cargo ecosystem.
\end{abstract}

\begin{IEEEkeywords}
rust, cargo, package manager, vulnerability propagation, knowledge graph
\end{IEEEkeywords}

%

\section{Introduction}
Rust is a statically typed programming language designed to improve performance and security by solving problems that C/C++ developers have long struggled with: memory errors and concurrent programming. Rust provides a way to import other libraries into your project, primarily through Rust's package manager, Cargo. These third-party libraries, known in the Rust ecosystem as crates, imported from the open source component central repository crates.io\cite{1cratesIo}, Cargo can help build code, download and compile third-party dependencies. Cargo is almost indispensable if you want to manage third-party dependencies when writing more complex Rust programs, but the Cargo ecosystem still faces problems with malicious dependencies\cite{2pfretzschner2017identification,3duan2020towards,4ohm2020towards,5gkortzis2021software,6ferreira2021containing,7prana2021out,8jafari2021dependency,9lauinger2018thou,zerouali2019impact}, vulnerability propagation\cite{10decan2018impact,11kikas2017structure,12liu2022demystifying}, poor compatibility\cite{8jafari2021dependency,13moller2020detecting,14hafner2021node}, license violations\cite{15decan2019package,16qiu2021empirical,decan2019empirical,zerouali2018empirical},technical Lag of dependencies\cite{cox2015measuring,decan2018evolution,chinthanet2019lag} and difficulty managing dependencies\cite{decan2016github,catuogno2017secure}.

Security vulnerabilities in open-source software third-party dependencies are one of the most pressing issues facing the open-source software supply chain; discovering and fixing vulnerabilities in packages can take a long time, and During this time, vulnerabilities can propagate to dependent packages, which poses a significant potential security risk to the open-source software supply chain. The December 2021 outbreak of the open-source vulnerability log4j\cite{log4j} is a stark example of these attacks, which exploit the increasing use of an open-source in the software development process, which is facilitated by dependency management that automatically parses, downloads, and installs hundreds of open source packages throughout the software lifecycle. The occurrence of such attacks.

Previous research has focused on vulnerability propagation in the ecosystem of Npm\cite{zimmermann2019small,6ferreira2021containing,8jafari2021dependency,cogo2021empirical}, RubyGems\cite{15decan2019package,11kikas2017structure,decan2017empirical}, Maven\cite{soto2021comprehensive,asyrofi2020ausearch}, Pypi\cite{decan2016topology,imminni2016spyse,valiev2018ecosystem,wang2020watchman}, packagist\cite{15decan2019package}, etc. Very little work has been done on vulnerability propagation in the Cargo ecosystem, In this study we focus on the security of dependencies in the Cargo ecosystem. Most scholars only consider direct dependencies in other studies on vulnerability propagation in ecosystems, and fewer scholars consider pass-through dependencies. However, they do not compare the pass-through dependencies with the actual official parsing rules, resulting in the limited accuracy of their research results. Our study not only considers direct dependencies and transmission dependencies but also takes into account the actual official parsing rules. Hence, it is more accurate and reasonable to study vulnerability propagation in the Cargo ecosystem through this approach. In this study, we focus on the security of dependencies in the Cargo ecosystem. We have conducted data mining and analysis of the official package registry of Cargo and combined the known security vulnerabilities of Rust\cite{GitHubAdvisoryDatabase} published on GitHub to build a knowledge graph of Cargo dependency vulnerabilities through the Neo4j\cite{Neo4j} graph database. rules and semantic\cite{Semantic} version control systems. Our results can help researchers better understand the vulnerability propagation problem in the Cargo package ecosystem, help the Rust community improve package review mechanisms, and reduce software supply chain attacks against the Rust language.
The challenges faced in this paper include the following:
\begin{itemize}
	\item {\verb|Data Analysis|}: The Cargo ecosystem's official package registry crates.io has released over 70,000 packages, each containing several different versions on average, each with different dependency information, forming a large and complex dependency network between dependencies, some of which may be discarded after release, making it extremely difficult to obtain, process, and analyze the data.
	
	\item{\verb|Building a knowledge graph|}: After obtaining data about the Cargo package ecosystem from crates.io, it is necessary to construct a knowledge graph based on the correlations between the data. The knowledge graph structure should be designed to accurately reflect the relationships between dependencies and provide data support for the subsequent dependency resolution algorithm, so it is difficult to design the structure of the knowledge graph of dependency vulnerabilities.
	\item{\verb|Dependency resolution algorithm design|}: The dependency resolution algorithm is the key to the accurate resolution of Cargo's dependency vulnerability knowledge graph to vulnerability propagation and is the key to determining the path of vulnerability propagation. The dependency resolution algorithm is challenging to satisfy both the official dependency resolution rules and to consider rules such as version control systems.
	\item{\verb|Propagation path determination|}: The dependency relationships in the Cargo package ecosystem are very complex, and it is difficult to build a good knowledge graph of dependency vulnerabilities and parsing algorithms to accurately calculate the actual path of vulnerability propagation.
\end{itemize}

The main contributions of this paper are as follows.
\begin{enumerate}
	\item [(1)] For the first time, a dependency vulnerability knowledge graph is constructed for the Cargo ecosystem, filling a gap in the field. We constructed the dependency-vulnerability knowledge graph containing 570563 nodes and 4023703 edges, covering all libraries and versions in the Cargo ecosystem in a specific time frame.
	\item [(2)] A new algorithm for parsing the dependency vulnerability knowledge graph is proposed, obtaining the dependency passing relationship consistent with the actual installation without installing the corresponding library. Our proposed parsing algorithm only needs to input the name and version number of the library to be parsed. According to the official dependency parsing rules, the algorithm can recursively calculate the dependency transfer relationship and save it as JSON data.
	\item [(3)] Based on the Cargo dependency-vulnerability knowledge graph and our proposed parsing algorithm, we conducted the first large-scale empirical study on vulnerability propagation paths, propagation scope, vulnerability characteristics, and vulnerability propagation factors in the Cargo ecosystem.Our study shows that the security vulnerabilities in the Cargo ecosystem are mainly memory-related. 18\% of the libraries affected by the vulnerability have their latest versions still affected the vulnerability. The number of versions affected by the vulnerability propagation in the whole Cargo ecosystem is 19.78\%. The percentage of libraries affected by vulnerability propagation in the whole Cargo ecosystem is 28.61\%.
	\item [(4)] Based on our findings, we propose feasible strategies that can be practically implemented to prevent the propagation of the vulnerability to the administrators of the Cargo community, the developers of the Cargo package manager, and the owners of the libraries respectively.
\end{enumerate}

The rest of the paper is organized as follows. Section II presents background information on the Cargo package management mechanism and its vulnerability propagation. Section III discusses the progress of existing work on other ecosystems and the Cargo ecosystem. Section IV presents the idea of constructing the Cargo dependency-vulnerability knowledge graph in this paper. Section V presents the design and implementation of the dependency-vulnerability knowledge graph parsing algorithm. Section VI presents the empirical study of vulnerability propagation in the Cargo ecosystem. Section VII describes the impact of this paper on the Cargo ecosystem and the Rust security community and the limitations of this paper. Section VIII summarizes the main contents of this paper.

\section{Motivation \& Background}
This section will describe Cargo's package management mechanism and some of the rules by which Cargo performs dependency resolution.This section also describes the problems with current approaches to studying the package management ecosystem, as well as why we chose the Knowledge Graph and why we wrote this article.

\subsection{Motivation}
Previously, the analysis of vulnerability propagation in the package manager ecosystem has mainly used static dependency analysis methods\cite{10decan2018impact,alfadel2021empirical}.If only the relationship between dependencies is statically resolved, this way of analyzing vulnerability propagation cannot accurately give the scope of vulnerability propagation in the package management system, because many library version constraints only give an upper limit, but not a lower limit, in this case we can hardly say that all versions below the upper limit will be affected by the vulnerability, so static dependency resolution has a great The limitations of static dependency resolution in calculating the scope of vulnerability propagation. A further problem with existing research is that it does not combine dependencies in the development environment with those in the production environment to analyze the actual propagation of vulnerabilities. In order to solve these problems, we propose a new dependency resolution algorithm by combining the official dependency resolution rules and static dependency resolution methods. Our proposed dependency vulnerability knowledge graph parsing algorithm can be more accurate than previous studies for vulnerability propagation analysis, and can better solve the problem of version range false positives caused by similar library version constraints when only an upper limit is given, because our algorithm can accurately calculate the list of versions that are specifically affected by the target vulnerability in the actual parsing process, instead of just giving a version range, which is our motivation for writing This is the motivation for writing this article.

\subsection{Why choose Knowledge Graph?}
Inspired by Liu et al\cite{12liu2022demystifying}, we find that knowledge graph is a good way to visualize information flow, which can better express the correlation between different versions of dependencies and vulnerabilities in the Cargo ecosystem, and at the same time, knowledge graph can connect structured and unstructured data in the Cargo ecosystem to solve the problem of information silos between these information sources, through which we We can connect libraries, different versions of libraries, and vulnerability data to form a vulnerability dependency knowledge graph that connects upstream and downstream dependencies, and then use this vulnerability dependency knowledge graph as a starting point to propose dependency resolution algorithms, which can better calculate the actual propagation of vulnerabilities in the current ecosystem, which is why we choose the knowledge graph as our research tool.

\subsection{Cargo dependency resolution rules}

Cargo allows users to specify the version of dependencies via the \emph{Cargo.toml} file. Let us take the example of quote\cite{quote}, a third-party package that has been downloaded 98 million times by the Rust community, to analyze how Cargo specifies the version of dependencies. If we specify quote = "1.0.16" in the \emph{Cargo.toml} file, where 1.0.16 appears to be a specific version number, but it represents a version range and allows compatibility updates under the constraints of the version control system Semantic, Table 1 analyzes Cargo's compatibility with the 1.0.16 version requirement as an example Table 1 analyzes the compatibility requirements of Cargo with version 1.0.16 as an example.

\begin{table}[]
	\setlength{\tabcolsep}{5ex} \centering
	\caption{Cargo version compatibility convention example analysis}
	\begin{tabular}{cc}
		\hline
		\textbf{Version Requirements} & \textbf{Version range}                    \\ \hline
		1.0.16                        & \textgreater{}= 1.0.16, \textless 2.0.0   \\
		1.0                           & \textgreater{}= 1.0.0, \textless{}2.0.0   \\
		1                             & \textgreater{}= 1.0.0, \textless{}2.0.0   \\
		0.0.16                        & \textgreater{}= 0.0.16, \textless{}0.0.17 \\
		0.0                           & \textgreater{}= 0.0.0, \textless{}0.1.0   \\
		0                             & \textgreater{}= 0.0.0, \textless{}1.0.0   \\ \hline
	\end{tabular}
\end{table}

Cargo uses Semantic to constrain the compatibility between different versions of a package. Cargo uses the leftmost non-zero number of the version to determine compatibility, e.g. version numbers 1.0.16 and 1.1.16 are considered compatible, and Cago considers it safe to update in the compatible range, but updates outside the compatibility range are not allowed. For example, updating from 1.0.16 to 2.0.0. Table 2 gives the syntax of Cargo's version requirements for dependencies.

\begin{table}[]
	\setlength{\tabcolsep}{3ex} \centering
	\caption{Cargo's version requirement syntax for dependencies}
	\label{tab:freq}
	\begin{tabular}{ccc}
		\hline
		\textbf{Symbols} & \textbf{Example}                  & \textbf{Version range}                 \\ \hline
		Insert           & \textasciicircum{}1.0.16          & \textgreater{}=1.0.16,\textless{}2.0.0 \\
		Wave             & $\sim$1.2                         & \textgreater{}=1.2.0,\textless{}1.3.0  \\
		Wildcard         & 1.*                               & \textgreater{}=1.0.0,\textless{}2.0.0  \\
		Equivalent       & =1.0.16                           & =1.0.16                                \\
		Compare          & \textgreater{}1.2                 & \textgreater{}=1.3.0                   \\
		Composite        & \textgreater{}=1.3,\textless{}1.5 & \textgreater{}1.3.0,\textless{}1.5.0   \\ \hline
	\end{tabular}
\end{table}

When Cargo encounters multiple packages specifying dependencies for a standard package, it first determines whether the versions of the dependencies specified by the multiple packages conform to the Semantic compatibility convention. It uses a giant version currently available in the compatibility range if they do. If they do not conform to the Semantic compatibility convention, Cargo builds two separate copies of the dependencies, but this may introduce a parsing error. Many of the versions in Cargo are pre-releases, which Cargo does not usually use. To use these pre-releases, the user must specify the pre-release version, which often means that it is unstable. The Semantic version requirement is not the only constraint considered by Cargo's dependency parser but also the characteristics of the package, the type of dependency, the version of the parser, and many other rules.

\section{Related work}

In this section, we will discuss security risks in the package ecosystem and work related to dependency resolution and management.

\subsection{Security risks in package ecosystem}
Many scholars have empirically analyzed the evolution of dependencies in package ecosystems over time, examining how existing package dependencies affect the ecosystem over time. Kikas et al.\cite{11kikas2017structure} analyzed the structure and evolution of dependency networks in JavaScript, Ruby, and Rust ecosystems, and their probing results revealed significant differences between language ecosystems. At the same time, their study shows that vulnerabilities in removing the most popular packages are increasing.

Li et al.\cite{li2022empirical} investigated how yanked values are used in the cargo ecosystem, as well as the reasons and frequency of use, and their findings show that from 2014 to 2020, the percentage of yanked use is increasing all the time, with package holders cancelling releases for other reasons than revoking flawed releases. They also found that 46\% of packages use delayed releases, which resulted in 1.4\% of releases in the ecosystem having unresolved dependencies.

Evans et al.\cite{evans2020rust} conducted a large-scale empirical study of the use of unsafe Rust in real-world Rust libraries and applications. Their study showed that the unsafe keyword is used in less than 30\% of Rust libraries and that more than half of Rust compilers cannot fully statically check for this problem because the unsafe keyword is hidden in the call chain the library. Bae et al.\cite{bae2021rudra} present a procedure for analyzing and reporting potential memory security vulnerabilities in unsafe Rust. They extend their analysis to all packages hosted in the Rust package registry, where RUDRA can scan the entire registry and identify unknown memory security vulnerabilities within 6.5 hours.

Decan et al.\cite{15decan2019package} empirically compare the compliance of Cargo, npm, Packagist, and Rubygems and examine the evolution of this compliance over time, exploring the extent to which ecosystem-specific features or policies affect the degree of compliance and presenting an assessment based on the principles of group wisdom to help package maintainers decide which type of versioning constraints they should impose on their dependencies.

Chinthanet et al.\cite{chinthanet2021lags} empirically investigated the fixed releases of packages from 231 npm projects on GitHub to determine the possible lag between vulnerable releases and their fixed releases, and their study lays the groundwork for how to mitigate lag in the ecosystem. Zerouali et al.\cite{zerouali2018empirical} proposed a lag model and validated it on the npm package manager This model, they analyzed the history of update times and technical lags for over 500,000 packages, considered development and runtime dependencies, and studied direct and pass-through dependencies.

\subsection{Dependency Analysis}

Liu et al.\cite{12liu2022demystifying} propose a knowledge graph-based dependency solution where they parse the dependencies in the Npm ecosystem into trees and investigate the security threats posed by dependency tree vulnerabilities on a large scale, and conduct an ecosystem-wide empirical study of vulnerability propagation in dependency trees and their evolution over time by precisely parsing the dependency trees with official dependency parsing rules.

Zimmermann et al.\cite{zimmermann2019small} study the security risks for npm users by systematically analyzing the dependencies between packages, the maintainers responsible for these packages, and publicly reported security issues, examining the possibility of running vulnerable and malicious code due to third-party dependencies. Their study finds that a single package may affect a large portion of the entire ecosystem, that a few maintainer accounts can inject malicious code into most packages, and that a lack of maintenance can cause packages to be vulnerable to attacks, while they give several mitigation techniques to face this problem.

Abate et al.\cite{abate2020dependency} review the idea of making dependency resolution a particular concern in package manager implementations, and by surveying the dependency resolution capabilities in state-of-the-art package managers, they argue that schemes such as SAT-based dependency resolution are being widely used, and present some new challenges for dependency resolution.

Catuogno et al.\cite{catuogno2017secure} addressed the problem of enforcing software dependencies in the package manager to prevent malicious users from forcing the system to install any package. At the same time, they performed an experimental evaluation of their protocol to update the critical material on the target device in a non-interactive manner. This critical update would allow decrypting more packages dependent on the new installation.

Pashchenko et al.\cite{pashchenko2018vulnerable} obtained a counting method to avoid overinflation by carefully analyzing deployed dependencies, aggregating dependencies by project, distinguishing stopped dependencies, addressing the overinflation of academic and industrial approaches to vulnerable dependencies in OSS software, and satisfying the need of industrial practices for proper allocation of development and audit resources, with the vast majority of vulnerable dependencies being able to be fixed by updating to the new version.

\section{Dependency-vulnerability knowledge graph construction}

In this section, we combine the characteristics of the dependency information in the Cargo software registry crates.io and the security vulnerabilities already disclosed in the Rust language to construct a Cargo dependency vulnerability knowledge graph to aid our subsequent vulnerability propagation research.


This paper relies on terms from the Rust software registry \emph{crates.io} and uses them in the graph database Neo4j, as described in Table 4.

To build the Cargo Dependency Vulnerability Knowledge Graph, we obtained the data of the library from the official Rust registry \emph{crates.io} and the disclosed vulnerabilities with CVE numbers from the GitHub Advisory. We then correlated these data according to specific dependencies through the graph database Neo4j.
\subsection{Dataset Source}
\textbf{Get library metadata.} We first monitored the GitHub repository corresponding to the official Rust software registry \emph{crates.io}, and obtained 75922 library metadata by cleaning and organizing the data with a Python script. The main field information of the metadata is shown in Table 5.

\textbf{Get the public CVE vulnerability of the Rust language.} We obtained the known vulnerabilities in the Rust language with CVE numbers disclosed on GitHub Advisory through a Python script and obtained a total of 351 vulnerabilities, the main fields contained in the vulnerabilities are shown in Table 6.

\textbf{Incremental updates of library metadata and CVE vulnerability data.}
\emph{crates.io}, the software registry corresponding to the Cargo package manager and the CVE vulnerability data corresponding to these libraries are constantly updated over time. To make our constructed knowledge graph of dependency-vulnerabilities effectively capture the updates of these data and keep our knowledge graph current, we used the requests library in Python to crates.io's data sources in the GitHub repository and the publicly available vulnerability data about the Rust language in the GitHub Advisory to The data is fetched every 2 hours and compared to the existing hash table. Our dataset is updated if there is an update to the data.

\subsection{Knowledge graph construction}
\textbf{The library metadata is associated with known CVE vulnerability data through the graph database Neo4j.} Where library, library\_version, and CVE vulnerability data are stored as nodes, where the edges include has, library\_affects, version\_affects, and version\_depends, etc. To better illustrate the connection between these nodes and relationships, we can refer to Figure 1, which describes the basic structure of Cargo's dependency-vulnerability knowledge graph.

\begin{figure}[h]
	\centering
	\includegraphics[width=1\linewidth]{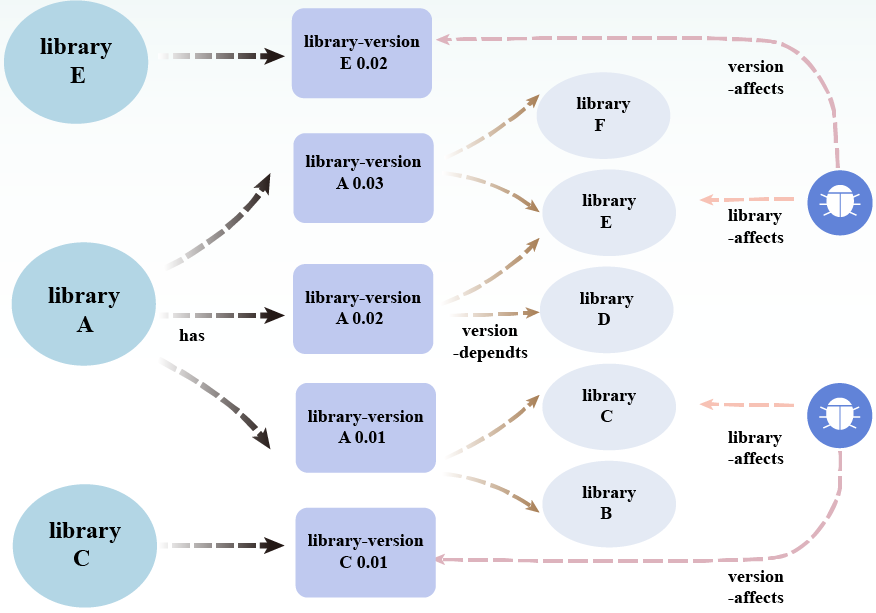}
	\caption{Cargo Dependency Vulnerability Knowledge Graph Structure Diagram}
\end{figure}

\textbf{Cargo Dependency-vulnerability Knowledge Graph data statistics.} In order to build an accurate Cargo dependency-vulnerability knowledge graph, we obtained 75922 library metadata from the official Rust software registry in January 2022 and 351 vulnerabilities in the Rust language that have been publicly disclosed and contain CVE numbers from the GitHub Advisory, of which there are library\_version nodes 494290. Among them are 4023698 Relationships, including 491744 has, 351 library\_affects, 6730 version\_affects and 3524878 version\_depends relationships, and the specific data can be referred to Table 3 Cargo dependency vulnerability knowledge graph data Statistics table.

\begin{table}[]
	\setlength{\tabcolsep}{5ex} \centering
	\caption{Cargo Dependency Vulnerability Knowledge Graph Statistics}
	\scalebox{1.2}{
		\begin{tabular}{cc}
			\hline
			\textbf{Nodes/Relationships} & \textbf{Statistics} \\ \hline
			library                      & 75922               \\
			cve                          & 351                 \\
			library\_version             & 494290              \\
			has                          & 491744              \\
			library\_affects             & 351                 \\
			version\_affects             & 6730                \\
			version\_depends             & 3524878             \\ \hline
		\end{tabular}
	}
	
\end{table}
\begin{table*}[]
	\setlength{\tabcolsep}{5ex} \centering
	\caption{Cargo Dependency Vulnerability Knowledge Graph Glossary}
	\label{tab:freq}
	\scalebox{1}{
		\begin{tabular}{ccc}
			\hline
			\textbf{Terminology} & \textbf{Description}                                                                                   & \textbf{Type} \\ \hline
			library              & Represents a separate software component in crates.io that can be referenced by other components.      & Node          \\
			library\_version     & Represents a certain version of a library.                                                             & Node          \\
			cve                  & Represents publicly disclosed vulnerabilities in the Rust language that have a CVE number.             & Node          \\
			has                  & library-\textgreater{}library\_version means that library has this version.                            & Relationship  \\
			library\_affects     & cve-\textgreater{}library represents a cve vulnerability that affects this library.                    & Relationship  \\
			version\_affects     & cve-\textgreater{}library\_version represents the cve vulnerability that affects the library\_version. & Relationship  \\
			version\_depends     & library\_version-\textgreater{}library represents the dependencies needed for a libray\_version.       & Relationship  \\ \hline
		\end{tabular}
	}
\end{table*}

\begin{table*}[]
	\setlength{\tabcolsep}{5ex} \centering
	\caption{ Example of library metadata field information (using abort as an example)}
	\begin{tabular}{ccc}
		\hline
		\textbf{Field Name}  & \textbf{Value}                         & \textbf{Description}                                  \\ \hline
		id                   & abort                                  & Database id                                           \\
		created\_at          & 2018-01-09T17:32:09.879845+00:00       & The time when the library was published to crates.io. \\
		description          & Abnormal termination (stable, no\_std) & Basic descriptive information about the library.      \\
		downloads            & 3506                                   & Total number of downloads                             \\
		max\_stable\_version & 0.1.3                                  & The most stable version number                        \\
		max\_version         & 0.1.3                                  & Maximum version number                                \\
		name                 & abort                                  & name                                                  \\
		newest\_version      & 0.1.3                                  & Latest version number.                                \\
		recent\_downloads    & 1972                                   & Number of recent downloads.                           \\
		updated\_at          & 2021-01-12T22:27:17.016095+00:00       & Last updated                                          \\ \hline
	\end{tabular}
\end{table*}

\begin{table*}[]
	\setlength{\tabcolsep}{5ex} \centering
	\caption{CVE-2022-21685 Vulnerability Field Information}
	\begin{tabular}{ccc}
		\hline
		\textbf{Field Name}    & \textbf{Value}                  & \textbf{Description}                            \\ \hline
		databaseId             & 9045                            & Description of the vulnerability in the CVE     \\
		severity               & MODERATE                      & Severity                                        \\
		cvss                   & 0.0                             & CVSS Scores                                     \\
		publishedAt            & 2022-01-14T21:03:36Z          & Release Time                                    \\
		summary                & Integer underflow in Frontier & Vulnerability Overview                          \\
		updatedAt              & 2022-01-15T00:03:46Z          & Update time                                     \\
		value                  & CVE-2022-21685                & Vulnerability Number                            \\
		vulnerableVersionRange & \textless{}= 0.1.0            & Range of versions affected by the vulnerability \\
		firstPatchedVersion    & null                            & First patch version                             \\
		ecosystem              & RUST                          & Ecosystem                                       \\
		package\_name          & frontier                      & Impacted package names                          \\ \hline
	\end{tabular}
\end{table*}

\section{Dependency-vulnerability knowledge graph parsing algorithm}

In this section, we design and implement a dependency-vulnerability knowledge graph parsing algorithm based on the Cargo dependency-vulnerability knowledge graph constructed above. In this algorithm, we consider the static analysis method and take into account the parsing rules of Cargo in actual operation to ensure the accuracy of our parsing algorithm as much as possible.

\subsection{Algorithm design}
Existing studies mainly study the dependency transfer relationships in ecosystems such as NPM and Maven. Most of them adopt a static analysis approach without considering the official resolution rules, resulting in a low accuracy rate of dependency transfer relationship resolution.

The hardware side of the algorithm experiment platform in this paper includes a Linux server with a 3090A GPU, and the software side includes the Neo4j graph database, Python v3.6.4, Py2neo v2021.2.3, and Semantic v2.13.0. Semantic is a library for implementing semantic version control that helps us determine the rule-compliant versions of a range of dependent versions. Neo4j is a high-performance NoSQL graph database that stores structured data on the web rather than in tables. Neo4j can also be seen as a high-performance graph engine with all the features of a mature database with the advantages of embeddedness, high performance, and lightweight.

Our parsing algorithm takes two parameters: the JSON metadata corresponding to the Cargo dependency-vulnerability knowledge graph and the package name and version number to be parsed, and the algorithm outputs JSON data containing the parent-child node relationship. The parsing algorithm finds the node data corresponding to the specified package name and version number and then iterates through all the dependencies required by this node. In considering the official parsing rules, we mainly consider whether it is a development dependency, whether the optional option is false, whether features are in the standard library, etc. At the same time, we have to determine the default version of the dependency by Semantic, pass the dependency by recursion, and finally, determine the hierarchical relationship based on the parent-child relationship between the dependencies and save the data in JSON format. The core of our algorithm is shown in Algorithm 1.

\begin{algorithm}
	\SetKwData{Left}{left}
	\SetKwData{This}{this}
	\SetKwData{LibraryNode}{LibraryNode}
	\SetKwData{DependencyData}{DependencyData}
	\SetKwData{CargoMap}{CargoMap}
	\SetKwData{Up}{up} 
	\SetKwData{Up}{up}
	\SetKwData{Up}{up}
	\SetKwFunction{Union}{Union}
	\SetKwFunction{OfficialParsingRules}{OfficialParsingRules}
	\SetKwFunction{DependencyResolution}{DependencyResolution}
	\SetKwData{ParsingList}{ParsingList}
	\SetKwFunction{FindNode}{FindNode} 
	\SetKwFunction{FindCompress}{FindCompress} 
	\SetKwInOut{Input}{input}
	\SetKwInOut{Output}{output} 
	
	\Input{Graph.json,Name of package,Version number} 
	\Output{Dependency tree metadata} 
	\BlankLine 
	
	\LibraryNode$\leftarrow$\FindNode{$Name,Version$}\;
	\ParsingList$\leftarrow$[]\;
	\CargoMap$\leftarrow$\{\}\;
	
	\SetKwProg{Fn}{Function}{:}{\KwRet}
	\Fn{\DependencyResolution{$Name,Version$}}{
		\For{$depend\leftarrow DependencyData[0]$ \KwTo $DependencyData[len-1]$}
		{ 
			\If{\OfficialParsingRules {$depend$} == True}{
				\ParsingList.append(depend)\;
				\CargoMap[name] $\leftarrow$ depend[version]\;
				\DependencyResolution(depend[name],depend[version])
			}\Else{continue}
		}
		
	}

	\SetKwProg{Fn}{Function}{:}{\KwRet}
	\Fn{\OfficialParsingRules{$depend$}}{
		\If{depend[type] == 'dev'}{
			\KwRet False\;
		}
		\If{depend[optional] == 'false'}{
			\KwRet False\;
		}
		\If{depend[features] not in depend['features']['std']}{
			\KwRet False\;
		}
		\Else{
			\KwRet True\;
		}
	}
	
	\caption{Knowledge Graph Parsing}
	
	\label{algo_disjdecomp} 
\end{algorithm}

\subsection{Algorithm Implementation}

We relied on the knowledge graph parsing algorithm constructed above to experiment with the top 100 downloaded packages in the official Rust package registry \emph{crates.io}, and achieved good parsing results. We will take the most downloaded package rand v0.8.5 of crates.io as an example to further illustrate the parsing results of our parsing algorithm.

By querying \emph{crates.io} we can see that the package \emph{rand} has been downloaded a total of 115,833,961 times since its release, and from the number of downloads, we can also see that this package is widely used. The result of our parsing algorithm is shown in Figure 2.

\begin{figure}[h]
	\centering
	\includegraphics[width=\linewidth]{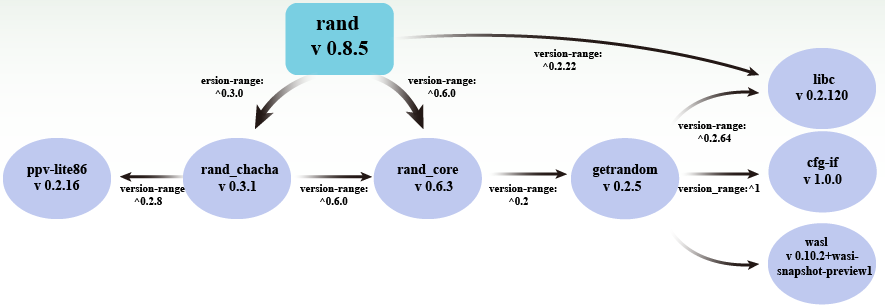}
	\caption{The result of the parsing algorithm for rand}
\end{figure}

Our parsing algorithm can parse the direct dependency of rand and the passed dependency and determine the version number that conforms to Cargo's official parsing rules based on the version range of the dependency. Maximum version number. We tested the actual installation by manually specifying the package name and version number in the \emph{Cargo.toml} file, and parsed the installed dependencies in the \emph{Cargo.lock} file with a Python script, and verified the validity of our algorithm by comparing a large amount of data.

\section{Vulnerability propagation study}

The dependency hierarchy between components in the Cargo ecosystem can lead to the propagation of vulnerabilities along the software supply chain. This section identifies the propagation paths of vulnerabilities in the Cargo ecosystem. It analyzes the possible propagation of vulnerabilities based on the Cargo dependency-vulnerability knowledge graph and parsing algorithm constructed above.

\subsection{Propagation path determination}

We take the library involved in the disclosed and CVE-numbered vulnerabilities in Cargo ecology as the starting point, combine it with our dependency tree parsing algorithm, reverse to find all the libraries containing this library in the dependency tree, and mark their paths and save them to the specified JSON file. We use this method for the 351 CVEs obtained vulnerabilities obtained by this method. In order to verify the accuracy of our vulnerability propagation paths, we also performed manual proofreading to ensure the consistency of the vulnerability propagation paths calculated by the Python program and the actual vulnerability propagation paths. We not only considered the static parsing rules when calculating the vulnerability propagation paths but also considered the official actual parsing rules to ensure the consistency between the experimental environment and the actual situation of vulnerability propagation.

To better illustrate the actual path of vulnerability propagation, we take the vulnerability CVE-2020-36442 as an example for explanation, as shown in Figure 3 CVE-2020-36442 vulnerability propagation path. This vulnerability affects the library beef in Rust. The range of versions affected by the vulnerability is <0.5.0, our parsing algorithm not only gives the specific range of versions affected beef also gives the actual list of affected versions while giving the vulnerability propagation path to pass the dependencies \emph{audiotags} and pass the dependencies on the actual list of affected versions The traditional static analysis method may calculate the library \emph{allaudiotags} in the vulnerability propagation path, but after our parsing algorithm found that \emph{allaudiotags} library depends on the 0.2.7182 version of \emph{audiotags}, but this version is not affected by the vulnerability, so In the Figure, we connect by dashed lines, the affected box in the Figure is the actual path of vulnerability propagation, so our algorithm can more accurately determine the actual path of vulnerability propagation, which provides an accurate data base for the following vulnerability propagation study.

\begin{figure}[h]
	\centering
	\includegraphics[width=\linewidth]{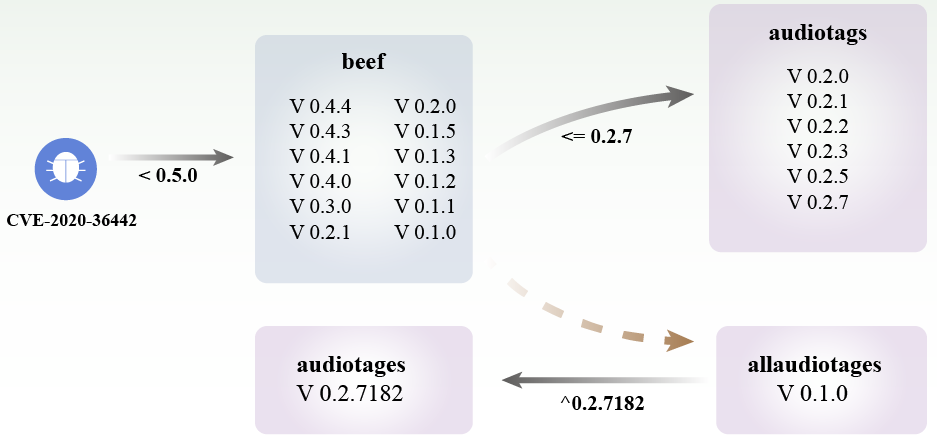}
	\caption{One of the propagation paths of CVE-2020-36442 vulnerability}
\end{figure}

\subsection{Statement of the Problem}
In order to accurately find the factors and characteristics of vulnerability propagation in the Cargo ecosystem, we conducted a large-scale evaluation of the libraries in crates.io based on the dependency-vulnerability knowledge graph parsing algorithm proposed in this paper. We launched a detailed investigation of the following issues.

\subsubsection{\textbf{RQ1: What are the characteristics of vulnerabilities in the Cargo ecosystem?}}
Using the publicly disclosed 351 vulnerabilities in the Rust language as a data base, we studied vulnerabilities in the Cargo ecosystem in terms of both the main types of vulnerabilities and the severity of the vulnerabilities. By studying the main types of vulnerabilities, we can reflect what kind of vulnerabilities the current Cargo ecosystem is mainly affected by, which can better warn developers using the Cargo ecosystem to avoid being affected by such vulnerabilities, and the severity of vulnerabilities can reflect the severity of vulnerability threats to libraries in the current Cargo ecosystem. The types of vulnerabilities described in this article are mainly from the main types given by the CWE\cite{CWE} community.

\textbf{The main types of vulnerabilities.} In order to understand the main types of vulnerabilities propagated in the Cargo ecosystem, we conducted vulnerability type statistics on the acquired vulnerabilities, and the statistics were compared using the CVE number as the standard using crawler technology and the NVD database, and the TOP 10 vulnerability types we counted are shown in Table 7 Cargo ecosystem vulnerability type TOP 10, from this table we can see that the vulnerability types affecting Cargo ecosystem security vulnerability types mainly include continued reference to memory after memory has been released, use of uninitialized resources, improper memory buffer boundary operations, double memory release, etc. It can be found that the current vulnerabilities affecting Cargo ecosystem security are mainly related to memory-related vulnerabilities, which is an interesting finding because we know that one of the features of the Rust language is This is an interesting finding because we know that one of the features of Rust language is that memory is safer compared to other languages, but we can find that most of the vulnerabilities in the Cargo ecosystem affect memory security, so users must be careful when managing and using third-party libraries through Cargo to prevent downloading libraries containing vulnerabilities through Cargo that could lead to memory security problems in the project.


\begin{table*}[]
	\setlength{\tabcolsep}{5ex} \centering
	\caption{Top 10 Cargo ecosystem vulnerability types}
	\scalebox{1}{\begin{tabular}{ccc}
			\hline
			\textbf{Type of vulnerability (CWE)} & \textbf{Description}                                                     & \textbf{Number of appearances} \\ \hline
			CWE-416                              & Use after release                                                        & 26                             \\
			CWE-908                              & Use of uninitialized resources                                           & 25                             \\
			CWE-119                              & Inappropriate restrictions on operations within memory buffer boundaries & 20                             \\
			CWE-415                              & Double Release                                                           & 19                             \\
			CWE-787                              & Cross-border memory writing                                              & 16                             \\
			CWE-362                              & Inappropriate concurrent execution of shared resources                   & 15                             \\
			CWE-77                               & Improper escape handling of special elements used in commands            & 14                             \\
			CWE-400                              & Uncontrolled resource consumption                                        & 9                              \\
			CWE-476                              & Null pointer dereference                                                 & 8                              \\
			CWE-125                              & Cross-border memory reading                                              & 8                              \\ \hline
	\end{tabular}}
\end{table*}

\textbf{Severity of vulnerabilities.} The severity of the 351 known vulnerabilities in the Cargo ecosystem was calculated from the GitHub Advisory, where the severity was classified as LOW (0.05), MODERATE (0.205), HIGH (0.460), and CRITICAL (0.328). The percentage of vulnerabilities classified as CRITICAL was 0.328, and the percentage of high-risk and urgent vulnerabilities was 0.788. The specific results are shown in Figure 4. We also analyzed the distribution of their CVSS scores. Our study found that the CVSS scores of vulnerabilities in the Cargo ecosystem were mainly distributed in the interval [4.7,9.8], as shown in Figure 5. (Cargo ecosystem vulnerability CVSS score distribution.)

\begin{figure}[h]
	\centering
	\includegraphics[width=1\linewidth]{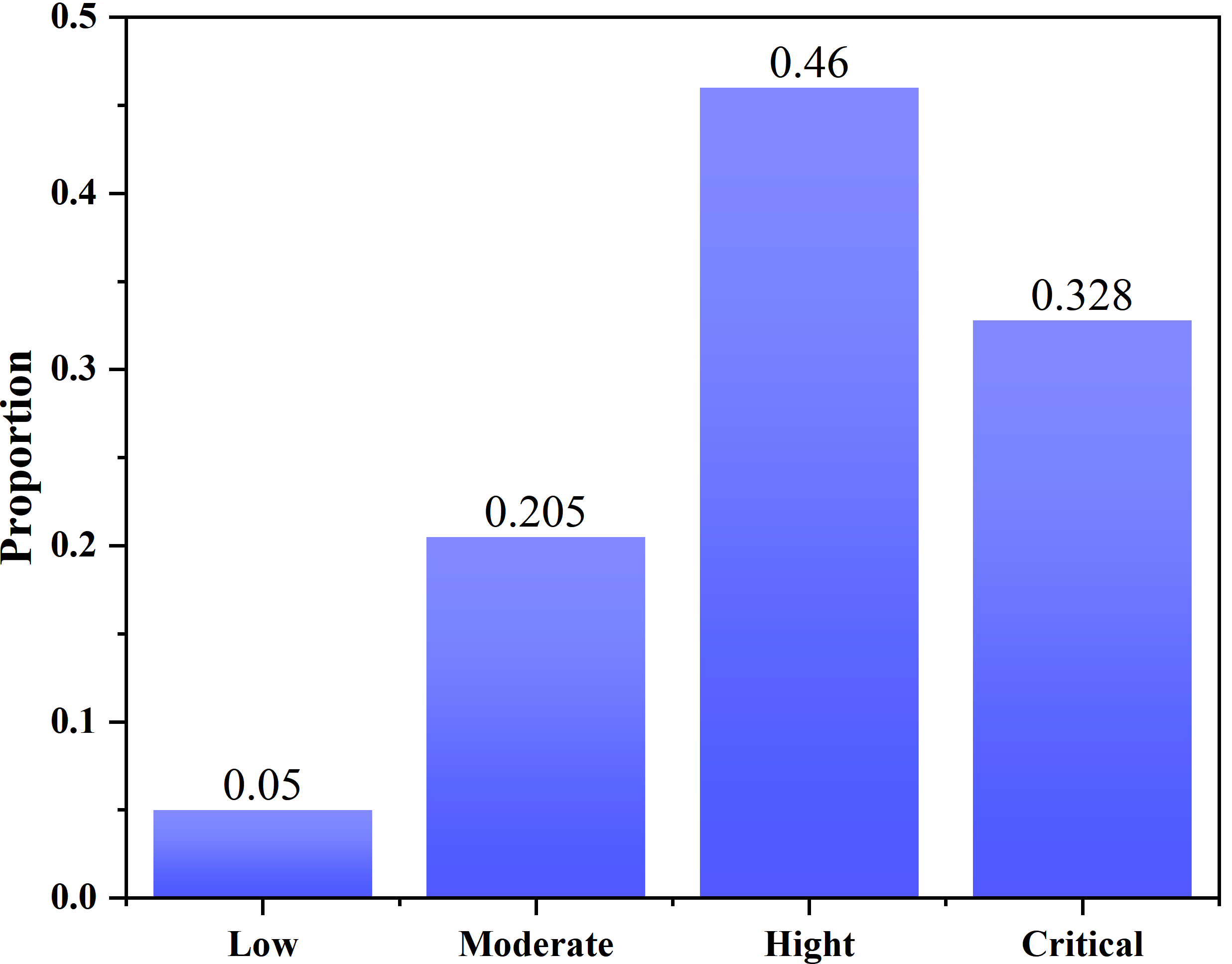}
	\caption{Cargo Ecosystem Vulnerability Severity Ratio Distribution}
\end{figure}

\begin{figure}[h]
	\centering
	\includegraphics[width=1\linewidth]{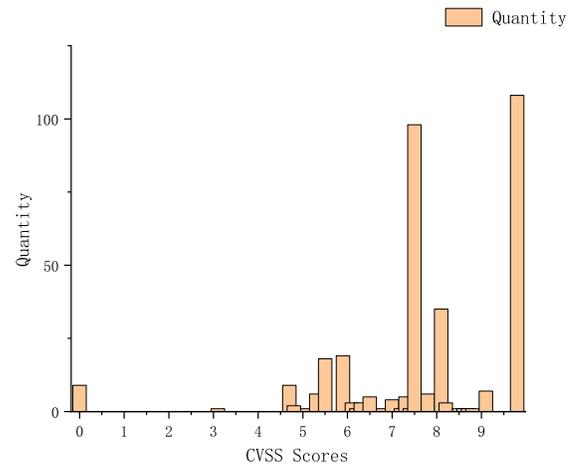}
	\caption{Cargo ecosystem vulnerability CVSS score distribution}
\end{figure}

\textbf{Proportion without patches.} We studied whether the vulnerabilities obtained had patches, and our study found that the Proportion of the 351 vulnerabilities obtained without patches accounted for 0.2678.We can see that the Proportion of vulnerabilities without patches in the Cargo ecosystem is very high. Once hacker implements targeted software supply chain attacks against these vulnerabilities may bring serious harm.

\subsubsection{\textbf{RQ2: Can vulnerable dependencies be fixed by updating the version?}}

We conducted a comparative study of the range of versions known to be affected by the vulnerability in the Cargo ecosystem and the latest versions of the affected libraries. We found that 82\% of the libraries affected by the vulnerability in the Cargo ecosystem could be avoided by updating to a new version, and 18\% of the latest versions of the affected libraries were still affected by the vulnerability. To better illustrate this type of problem, we selected five instances where the vulnerabilities have been published for years. However, the latest version of the library is still affected by the vulnerabilities, as shown in Table 8. Our statistics found that many of the latest versions of libraries in the Cargo ecosystem are still affected by the vulnerabilities. These vulnerabilities can be downloaded and used by users through cargo, which brings the possibility of software supply chain attacks.
\begin{table*}[]
	\setlength{\tabcolsep}{3ex} \centering
	\caption{Example of the latest version of library still affected by the vulnerability}
	\scalebox{1}{
		\begin{tabular}{ccccccc}
			\hline
			\textbf{CVE-ID} & \textbf{Affected library} & \textbf{Range of versions affected} & \textbf{Latest version} & \textbf{Vulnerability Announcement Time} \\ \hline
			CVE-2016-10933  & portaudio                 & \textless{}=0.7.0                   & 0.7.0                   & 2019-8-26                                                     \\
			CVE-2020-35900  & array-queue               & 0.3.3                               & 0.3.3                   & 2020-12-31                                                   \\
			CVE-2021-30456  & id-map                    & 0.2.1                               & 0.2.1                   & 2021-4-7                                                  \\
			CVE-2021-29936  & adtensor                  & \textless{}=0.0.3                   & 0.0.3                   & 2021-4-1                                                       \\
			CVE-2020-36204  & im                        & \textless{}=15.0.0                  & 15.0.0                  & 2021-1-26                                                   \\ \hline
		\end{tabular}
	}
	
\end{table*}

\subsubsection{\textbf{RQ3: What is the scope of vulnerability propagation?}}
This paper studies the vulnerability propagation of 351 known vulnerabilities in the Cargo ecosystem. Based on our proposed dependency vulnerability knowledge graph parsing algorithm and vulnerability propagation path determination method above, after our calculations, we find that there are 246 libraries containing known vulnerabilities. A total of 6731 versions of these libraries are affected by the vulnerabilities. There are 21722 libraries affected by the propagation of pass-dependent vulnerabilities, of which a total of 97,779 versions of these libraries are affected by the propagation of vulnerabilities, which indicates that the percentage of libraries affected by the propagation of vulnerabilities in the entire Cargo ecosystem is 28.61\% (21722/75922). The percentage of versions affected by vulnerabilities in the entire Cargo ecosystem is 19.78\% (97779/494290). From this percentage, we can see that the impact of vulnerability propagation in the Cargo ecosystem is not negligible.

We counted the five most serious CVE vulnerabilities affecting the security of the Cargo ecosystem, including the libraries directly affected by these vulnerabilities, as well as the library affected by the propagation of the vulnerabilities and the total number of versions affected. This is shown in Table 9.
\begin{table*}[]
	\setlength{\tabcolsep}{3ex} \centering
	\caption{The five most serious CVE vulnerabilities affecting the security of the Cargo ecosystem}
	\begin{tabular}{cccc}
		\hline
		\textbf{CVE-ID} & \textbf{Direct impact library} & \textbf{Total number of libraries spreading} & \textbf{Total number of versions influenced} \\ \hline
		CVE-2021-32715  & hyper                          & 2592                                         & 19898                                        \\
		CVE-2021-45710  & tokio                          & 3630                                         & 18461                                        \\
		CVE-2019-25010  & failure                        & 3252                                         & 16923                                        \\
		CVE-2018-20997  & openssl                        & 622                                          & 5144                                         \\
		CVE-2020-35916  & image                          & 876                                          & 5066                                         \\ \hline
	\end{tabular}
\end{table*}

Our research has found that a known CVE vulnerability may affect only one library during public disclosure. However, this library's propagation scope in the open-source software supply chain can be enormous, and the impact on the library through propagation may be overwhelming. Let us imagine a scenario where a hacker manages to obtain credentials to push a malicious version of a library to a location in the famous \emph{crates.io} dependency tree. Then all users of the popular library are affected by this vulnerability once the malicious library is involved in the dependency tree of the famous library.

\subsubsection{\textbf{RQ4: Is the version containing the vulnerability being deprecated?}}
The yank mechanism is a mechanism that Cargo provides to developers to deprecate a library that has been published to the software registry crates.io. The package manager provides a deprecation mechanism to developers to organize the use of certain features in the library, such as API methods. However, in our research, we found that the percentage of libraries whose latest versions are still affected by vulnerabilities and whose latest versions are yanked to true is only 1.7\%, which means that a high percentage of libraries in the Cargo ecosystem are not deprecated even though the latest versions have security vulnerabilities. This also shows that there is still a shortage of release level deprecation of libraries with vulnerabilities in the Cargo ecosystem.

\subsubsection{\textbf{RQ5: What are the main factors that cause the propagation of vulnerabilities in the Cargo ecosystem?}}

In this paper, the vulnerability propagation study in the Cargo dependency vulnerability knowledge graph found that the main factors causing the propagation of vulnerabilities in the Cargo ecosystem are the following.

\textbf{Ignore the impact of passing dependencies.} Many developers only focus on the security of the packages they download and parse through Cargo but ignore the security of the pass-through dependencies that come with the packages. The security impact caused by direct dependencies is far less than the security risk caused by pass-through dependencies. We found that the spread of vulnerabilities caused by pass-through dependencies is about ten times greater than direct dependencies when determining the spread of vulnerabilities.

\textbf{Not updated to the latest version.} After an existing vulnerability is discovered, developers are often notified to fix the vulnerability first, and it may only be disclosed through public vulnerability databases such as NVD after a specified period, so responsible developers tend to address the existence of security vulnerabilities by updating the version, so a large percentage of the vulnerability propagation in the Cargo ecosystem is caused by not updating the version of the library containing the vulnerability promptly. Library.

\textbf{Libraries with known vulnerabilities are not being addressed.} Our research found that 60 of the latest versions of crates.io's libraries were affected by security vulnerabilities that have been published for several years. However, the crates.io community did not deprecate or take other measures to address these vulnerable libraries at the release level, nor did they receive any warning messages when installing them through Cargo. Therefore, the failure of the community maintainers to effectively deal with these vulnerable libraries is an essential factor in the spread of the vulnerability.

\section{Discussion}

In this section, we discuss our findings and the implications of these findings for the Rust language, library maintainers, and the Cargo community. At the same time, we give measures on how to mitigate these software supply chain attacks against the Cargo ecosystem for the characteristics of vulnerability propagation in the Cargo ecosystem, and finally, conclude with the limitations of this paper.

\subsection{Impact}
We look at the factors that contribute to the propagation of vulnerabilities in the Cargo ecosystem and hope to provide some valuable insights for managers, users, and owners of libraries in the Cargo community while contributing to the future development of the Cargo ecosystem.

\textbf{Advice to the administrators of the Cargo community.} Monitor in real-time the vulnerability information disclosed by the mainstream security vulnerability database about the existing library on crates.io, ensure as much as possible that the vulnerability data release is aware, and notify the maintenance staff of the library to deal with the library, release the version that fixes the vulnerability as early as possible, and before the developers release the version that fixes the vulnerability, once a user has downloaded this library through cargo to download this library, immediately give a vulnerability warning message to remind users that the library being downloaded contains known vulnerabilities, and users who have already downloaded this library in the project can be reminded by email. If the library owner fails to fix the vulnerability within a specified period, a mechanism can take the library down. At the same time, we note that developers in the Rust community have proposed a code review system for the Cargo package manager, \emph{cargo-crev}, a tool that helps Rust users assess the quality and trustworthiness of their package dependencies, so we recommend that the Cargo community increase support for these tools.

\textbf{Advice to Cargo users.} When installing a library through Cargo, the user should not only confirm whether this library contains known CVE vulnerabilities but also check whether the delivery dependency network of this library contains known vulnerabilities through the \emph{Cargo.lock} file as much as possible, and if the project is enormous. It is not easy to find it manually. Ensure the installed library is The latest version because the commonly used library will generally be upgraded to the fixed version before the vulnerability is publicly disclosed.
Meanwhile, Cargo officially provides the Cargo-audit tool to help users check for vulnerabilities in the RustSec vulnerability database. Users can use this tool to check the libraries involved in the \emph{Cargo.lock} file in their projects to prevent possible vulnerabilities.

\textbf{Advice to library owners.} If the owner receives a bug or vulnerability message from the crates.io community or open-source repository, the owner should be able to locate the scope of the vulnerability quickly and fix it and release a new version to \emph{crates.io} as soon as possible. Suppose the vulnerability cannot be fixed in the short term. In that case, the owner can also contact the administrators of the Cargo community to assist with the process of preventing software supply chain attacks through these vulnerabilities.

\subsection{Limitations}
Due to the limitation of computing power, we only calculated the general situation of vulnerability propagation. We did not make a deeper calculation of the nodes involved in more vulnerability propagation paths. However, the case of vulnerability for multi-level propagation in our experiments is relatively small. Our study only ignored the version of the propagation path of more than 100. This case of vulnerability propagation only accounts for 1.78\% of the total number of experiments. Therefore, the impact on the overall experimental results is relatively small, and the impact on our experimental conclusions is negligible.

The vulnerability data may not fully cover the existing vulnerabilities in the Cargo ecosystem. However, the existing vulnerabilities indicate the problems in the Cargo ecosystem, so the impact on the overall findings of the study is relatively small.

In future work, we will increase the arithmetic power and optimize the time complexity of our algorithm to investigate the case of multi-level propagation of vulnerabilities. We will also collect more vulnerabilities from NVD, CVE, RUSTSEC and other vulnerability databases about the Cargo ecosystem for further research.

\section{Conclusion}
In order to empirically study the propagation of vulnerabilities in the Cargo ecosystem
In this paper, we fill the gap in the field by constructing a Cargo dependency-vulnerability knowledge graph containing 570,563 nodes and 4,023,703 edges based on the Cargo dependency parsing rules and 75,922 known security vulnerabilities of libraries and Rust language in the Cargo ecosystem as the data base. In this paper, from the constructed knowledge graph, we propose for the first time a dependency-vulnerability knowledge graph parsing algorithm considering the official parsing rules of Cargo, which can accurately simulate the actual dependency passing of a given library and version number without actually installing the library. Based on this algorithm, we study the vulnerability propagation problem in the Cargo ecosystem and analyze the path of vulnerability propagation. We give the scope of vulnerability propagation in the Cargo ecosystem and the factors that may lead to vulnerability propagation. Finally, we propose some measures for the Cargo community to mitigate these problems.


\bibliographystyle{IEEEtran}
\bibliography{sample-base}
\end{document}